\documentclass[preprint,floats,nofootinbib,superscriptaddress,color,showpacs]{revtex4}

\usepackage{bm}
\usepackage{amsmath}
\usepackage{graphicx}
\usepackage{graphics}
\usepackage[dvips]{color}

\begin{document}

\title{Relativistic analyses of quasielastic neutrino cross sections at MiniBooNE kinematics}
\author{J.E. Amaro}
\affiliation{Departamento de F\'{\i}sica At\'{o}mica, Molecular y Nuclear,
Universidad de Granada,
  18071 Granada, SPAIN}
\author{M.B. Barbaro}
\affiliation{Dipartimento di Fisica Teorica, Universit\`a di Torino and
  INFN, Sezione di Torino, Via P. Giuria 1, 10125 Torino, ITALY}
\author{J.A. Caballero}
\affiliation{Departamento de F\'{\i}sica At\'{o}mica, Molecular y Nuclear,
Universidad de Sevilla,
  41080 Sevilla, SPAIN}
\author{T.W. Donnelly}
\affiliation{Center for Theoretical Physics, Laboratory for Nuclear
  Science and Department of Physics, Massachusetts Institute of Technology,
  Cambridge, MA 02139, USA}
\author{J.M. Ud\'{\i}as}
\affiliation{Grupo de F\'{\i}isica Nuclear, Departamento de
F\'{\i}sica At\'{o}mica, Molecular y Nuclear,
Universidad Complutense de Madrid,
  28040 Madrid, SPAIN}

\begin{abstract}
Two relativistic approaches are considered to evaluate the
quasielastic double-differential and integrated neutrino-nucleus
cross sections. One, based on the relativistic impulse
approximation, relies on the microscopic description of nuclear
dynamics using relativistic mean field theory, and incorporates a
description of the final-state interactions. The second is based
on the superscaling behavior exhibited by electron scattering data
and its applicability, due to the universal character of the
scaling function, to the analysis of neutrino scattering
reactions. The role played by the vector meson-exchange currents
in the two-particle two-hole sector is also incorporated and the
results obtained are compared with the recent data for neutrinos
measured by the MiniBooNE Collaboration.
\end{abstract}


\maketitle

\section{Introduction}
\label{sec:intro}

The data on muon neutrino charged-current quasielastic (CCQE)
cross sections recently obtained by the MiniBooNE
collaboration~\cite{AguilarArevalo:2010zc}, and its comparison
with several theoretical calculations, have led to an important
debate concerning the role played by various ingredients entering
in the description of the reaction: nuclear dynamics (final-state
interactions (FSI), low-lying nuclear excitations, effects beyond
the impulse approximation (IA), {\it etc.}), as well as possible
modifications of the single-nucleon form factors. Although no
definitive conclusions are yet in hand, a detailed study of
modeling versus experiment for inclusive quasielastic electron
scattering and its extension to neutrino processes can shed light
on the different interpretations of the discrepancy between theory
and experiment.

When a dipole shape is assumed for the axial form factor, the
nucleon axial mass $M_A$ can be considered to be the only free
parameter within the Relativistic Fermi Gas (RFG) model, presently
used in many Monte Carlo codes employed in the analysis of the
experimental data. When compared with MiniBoone CCQE data,  the RFG
underestimates the total cross section unless an axial mass $M_A$ of
the order of 1.35 GeV/c$^2$ is employed in the dipole prescription
for the form factor. This value of the axial mass is considerably
larger than the accepted world average value \cite{Bernard:2001rs},
thus yielding a larger axial form factor. This should be taken more
as an indication of incompleteness of the theoretical description of
the MiniBooNE data based upon the RFG, rather than as a true
indication for a larger axial mass.

For instance, although the RFG incorporates a fully relativistic
treatment, required by the kinematics of the experiment (mean
neutrino energy flux, $\langle E_\nu \rangle=788$ MeV, with values
up to 3 GeV), its description of the nuclear dynamics is clearly
too crude to draw specific conclusions on the value of the
anomalous axial mass from the departure of the RFG from
experiment, but rather as a hint on the importance of nuclear
effects in describing these experimental data.

However, at the level of the impulse approximation, a number of
much more sophisticated descriptions of the nuclear dynamics other
than the one represented by the RFG, based for instance on the use
of realistic spectral functions \cite{Barbaro:1996vd,Ben10,Jus10},
when compared with the MiniBooNE experimental data also
underpredict the measured CCQE cross section, in this respect not
doing a better job than the RFG. One important consideration that
must be taken into account is the fact that, even when these
models provide a much more realistic description of the nuclear
dynamics than the RFG, they are built on non-relativistic
approaches that are likely questionable at the kinematics of the
MiniBooNE experiment.

Among the difficulties that one faces when comparing models, is that
the effect of the ingredients in the model, such as interactions in
the final state, may differ greatly from model to model. For example
in~\cite{Ben10} the FSI are barely seen, causing only a simple
$\sim$10~MeV shift of the QE peak. However, relativistic and
semi-relativistic models  of inclusive QE $(e,e')$ reactions which
included a relativistic mean field, that is, described the FSI by
means of strong relativistic potentials or their semi-relativistic
equivalents, have clearly shown the essential role played by FSI in
order to describe properly the behavior of
data~\cite{Maieron:2003df,Caballero:2006wi,Amaro:2006if,Caballero:2009sn,Meucci:2009nm}.

In addition to the relativistic treatment of the nuclear
excitations, in some regions of the wide range of neutrino energies
where the neutrino flux for the experiment has significant strength,
the reaction may have sizable contributions from effects beyond the
IA. For instance, in \cite{Mar09,Mar10} when the theoretical results
incorporated multiple knockout excitations, they were shown to be in
accordance with the total cross section data without the need to
increase the value of $M_A$. However, no comparison with the
experimental double-differential cross section is shown in
\cite{Mar09,Mar10}. Moreover, these calculations are based on
non-relativistic reductions whose reliability at MiniBooNE
kinematics may be doubtful. In fact, the kinematics of the MiniBooNE
experiment demands relativity as an essential ingredient; not only
relativistic kinematics should be considered, but also the nuclear
dynamics and current operators should be described within a
relativistic framework \cite{Amaro:1998ta,Amaro:2002mj}.
Furthermore, the wide range of neutrino energies, at least for some
specific conditions, may also require one to account for effects not
included in models devised for quasi-free scattering. This is, for
instance, the situation at the most forward scattering angles where
a very significant contribution in the cross section may come from
very low-lying excitations in nuclei~\cite{Amaro11}.

A systematic analysis of the world inclusive $(e,e')$ data has
clearly demonstrated that, for sufficiently large momentum
transfers, at energy transfers below the QE peak the property of
superscaling works rather
well~\cite{Day:1990mf,Donnelly:1998xg,Donnelly:1999sw,Maieron:2001it},
that is, the reduced cross section, when represented versus the
scaling variable~\cite{Alberico:1988bv}, is largely independent of
the momentum transfer (first-kind scaling) and of the nuclear target
(second-kind scaling). Moreover, from the longitudinal response a
phenomenological scaling function has been extracted that shows a
clear asymmetry with respect to the QEP with a long tail extended to
positive values of the scaling variable (larger energy transfers).
Assuming the scaling function to be universal, {\it i.e.,} valid for
electromagnetic and weak interactions,
in~\cite{Amaro:2004bs,Amaro:2005dn} CCQE neutrino-nucleus cross
sections were evaluated by using the scaling function extracted from
$(e,e')$ data and multiplying it by the corresponding elementary
weak cross section. This approach, denoted simply as ``SuSA'',
provides nuclear-model-independent neutrino-nucleus cross sections,
but its reliability rests on a basic assumption: the scaling
function (extracted from longitudinal $(e,e')$ data) is appropriate
for all of the various weak responses involved in neutrino
scattering (charge-charge, charge-longitudinal,
longitudinal-longitudinal, transverse and axial), and is independent
of the vector or axial nature of the nuclear current entering the
hadronic tensor. In particular, the SuSA approach assumes the
electromagnetic longitudinal (L) and transverse (T) scaling
functions to be equal. This property, known as scaling of the zeroth
kind, is fulfilled by the RFG (by construction) and by most models
based on non-relativistic descriptions that, in a way, factorize the
elementary lepton-nucleus amplitude into a lepton-nucleon part and a
part containing the nuclear effects
\cite{Vignote:2003er,Udias:1993zs}. Within SuSA, this factorization
in the elementary amplitude propagates even to the cross section,
which is then proportional to the elementary lepton-nucleon cross
section and to the nuclear response, the latter in this approach
being a universal function.

However, from the analysis of the existing L/T separated data, after
removing inelastic contributions and two-particle-emission effects
one finds that the ``purely nucleonic'' transverse scaling function
is significantly larger than the longitudinal one~\cite{private11}.
This has to be attributed to a breakdown of the elementary
factorization mentioned before, so that the elementary
lepton-nucleon vertex inside the nucleus is no longer accurately
described by the one for free nucleons. One must resort to models
such as the relativistic mean field approach, denoted as RMF, where
the relativistic dynamics introduces significant deviations of the
behaviour of the elementary lepton-nucleon vertex in the presence of
strong scalar and vector potentials \cite{Vignote:2003er}. This
breakdown of zeroth-kind scaling present in the RMF seems to be
favored by the comparison with data~\cite{private11}.

In a recent paper~\cite{Amaro11} SuSA predictions have been compared
with the MiniBooNE data for the double-differential neutrino cross
section showing a systematic discrepancy between theory and
experiment. Inclusion of 2p-2h Meson Exchange Current (MEC)
contributions yields larger cross sections and accordingly better
agreement with the data. However, theory still lies below the data
at larger angles where the cross sections are smaller. Before
drawing definitive conclusions on the anomalous axial mass, it is
important to explore alternative approaches that have been shown to
be successful in describing inclusive QE $(e,e')$ processes. As just
mentioned, this is the case for the RMF, where a fully relativistic
description (kinematics and dynamics) of the process is
incorporated, and FSI are taken into account by using the same
relativistic scalar and vector energy-independent potentials
considered in the description of the initial bound states. The RMF
model applied to inclusive QE $(e,e')$ processes has been shown to
describe scaling behaviour, and more importantly, it gives rise to a
superscaling function with a significant asymmetry, namely, in
complete accord with data~\cite{Caballero:2006wi,Caballero:2007tz}.
Moreover, contrary to SuSA, where scaling of the zeroth kind is
assumed, the RMF model provides longitudinal and transverse scaling
functions which differ by typically $20\%$, the T one being larger.

The RMF approach has been applied to the description of CCQE
neutrino-nucleus cross sections~\cite{Caballero:2005sj,Amaro:2006tf,Maieron:2003df,Martinez:2005xe}
and it has been investigated with respect to how scaling emerges
from neutrino reactions, and how the {\sl ``theoretical''}
neutrino scaling functions compare with the corresponding ones
evaluated for electrons (L and T responses) and with the data
\cite{Caballero:2005sj,Amaro:2006tf}.

Even at the level of the impulse approximation, the zeroth-kind
scaling violation introduced by the RMF approach, as well as the
different isospin character shown by the electromagnetic and weak
nucleon form factors, can lead to significant discrepancies between
the results provided by SuSA and RMF approaches. Furthermore,
effects beyond the IA give rise to additional scaling violations in
the transverse responses. Thus, a proper relativistic description of
these effects is needed in order to compare with the data taken by
the MiniBooNE collaboration.

The paper is organized as follows: after this introductory section, in the one following (Sect.~\ref{sec:analysis}) we present an analysis of the results obtained using three models for the CCQE cross sections, while in Sect.~\ref{sec:concl} we end by making a few concluding remarks.

\section{Analysis of results}
\label{sec:analysis}

In this section we discuss the results obtained with the different approaches considered and compare with the experimental data. Details on the various approaches considered have been presented in previous works. In particular, the SuSA approach and its extension to CC neutrino reactions can be reviewed in
\cite{Amaro:2004bs}, whereas the basic ingredients entering in the RMF model applied to inclusive electron and CCQE neutrino
reactions are given in \cite{Caballero:2006wi,Caballero:2007tz,Caballero:2005sj,Amaro:2006tf,Maieron:2003df,HerraizZZ}.

We first show results for the CCQE $\nu_\mu$--$^{12}$C
double-differential cross section averaged over the neutrino flux
$\Phi(E_\nu)$, namely
\begin{equation}
\frac{d^2\sigma}{dT_\mu d\cos\theta}
= \frac{1}{\Phi_{tot}} \int
\left[ \frac{d^2\sigma}{dT_\mu d\cos\theta} \right]_{E_\nu}
\Phi(E_\nu) dE_\nu ,
\end{equation}
where $T_\mu$ and $\theta$ are the kinetic energy and scattering
angle of the outgoing muon, $E_\nu$ is the neutrino energy and
$\Phi_{tot}$ is the total flux. For each value of the neutrino
energy the above cross section can be expressed in terms of seven
nuclear response functions as \cite{Amaro:2004bs}
\begin{eqnarray}
\left[ \frac{d^2\sigma}{dT_\mu d\cos\theta} \right]_{E_\nu} &=&
\sigma_0 \left[ {\hat V}_L R_L^{VV} + {\hat V}_{CC} R_{CC}^{AA} +
2{\hat V}_{CL} R_{CL}^{AA} + {\hat V}_{LL} R_{LL}^{AA} \right.
\nonumber
\\
&&\qquad\qquad \left. + {\hat V}_T \left( R_T^{VV} + R_T^{AA}
\right) + 2 {\hat V}_{T^\prime} R_{T^\prime}^{VA} \right] ,
\end{eqnarray}
where ${\hat V}_i$ are kinematic factors and the indices
$L,C,T,T^\prime,V,A$ refer to longitudinal, charge, transverse,
transverse-axial, vector and axial-vector components of the
nuclear current, respectively.

In particular, in the SuSA approach each response function can be
cast as
\begin{equation}
R_i (q,\omega) = \frac{m_N}{q k_F} R_i^{sn}(q,\omega) f(\psi) ~,
\end{equation}
where $q$ qnd $\omega$ are the transferred momentum and energy,
respectively, $m_N$ is the nucleon mass, $k_F$ is the Fermi
momentum, $R_i^{sn}$ are the single-nucleon responses,
$\psi(q,\omega)$ is the RFG scaling variable (see, {\it e.g.},
\cite{Alberico:1988bv} for its definition) and $f(\psi)$ is the
so-called superscaling function, containing the dependence on the
nuclear model. In the SuSA model it is given by a fit to the
experimental longitudinal $(e,e^\prime)$ reduced response
function~\cite{Jourdan:1996ut}.

In the RMF case, the weak response functions are given by taking
the appropriate components of the weak hadronic tensor constructed from the single-nucleon current:
\begin{equation}
\langle J_W^\mu\rangle = \int d{\bf r}\overline{\phi}_F({\bf r})\hat{J}_W^\mu({\bf r})\phi_B({\bf r})
\, ,
\end{equation}
where $\phi_B$ and $\phi_F$ are relativistic bound-state and
scattering wave functions, respectively, and $\hat{J}_W^\mu$ is
the relativistic one-body current operator modeling the coupling
between the virtual $W$ and a nucleon~\cite{Coulomb}. The bound
nucleon states are described as self-consistent Dirac-Hartree
solutions, derived within an RMF approach by using a Lagrangian
containing $\sigma$, $\omega$ and $\rho$
mesons~\cite{Horowitz,Serot,Sharma:1993it}. The outgoing nucleon
wave function is computed by using the same relativistic mean
field employed in the initial state. This incorporates the FSI
between the ejected nucleon (proton) and the residual nucleus.

Finally, concerning the description of MEC contributions, we use
the fully relativistic model of
\cite{Amaro:2002mj,Amaro:2003yd,Amaro:2009dd}. In particular in
the 2p-2h sector we use the scheme applied in \cite{De
Pace:2003xu} to electron scattering, where all many-body diagrams
containing two pionic lines were taken into account. However, it
is important to point out that, within the present approach, only
the pure vector transverse response, $R_T^{VV}$, is affected by
MEC. Effects in the axial-vector transverse response, as well as
the contribution of the correlation diagrams (performed recently
for $(e,e')$ in \cite{Amaro:2p2h}), should be incorporated into
the analysis before definitive conclusions on the comparison with
data can be drawn. Work along this line is presently under way.
MEC contributions to $\nu$--$^{12}$C reactions have been computed
within a somewhat different approach both for charged and neutral
currents, in \cite{Umino:1996cz,Umino:1994wu}, where the effect of MEC in the cross section was found to be less than 10\%.

\begin{figure*}[ht]
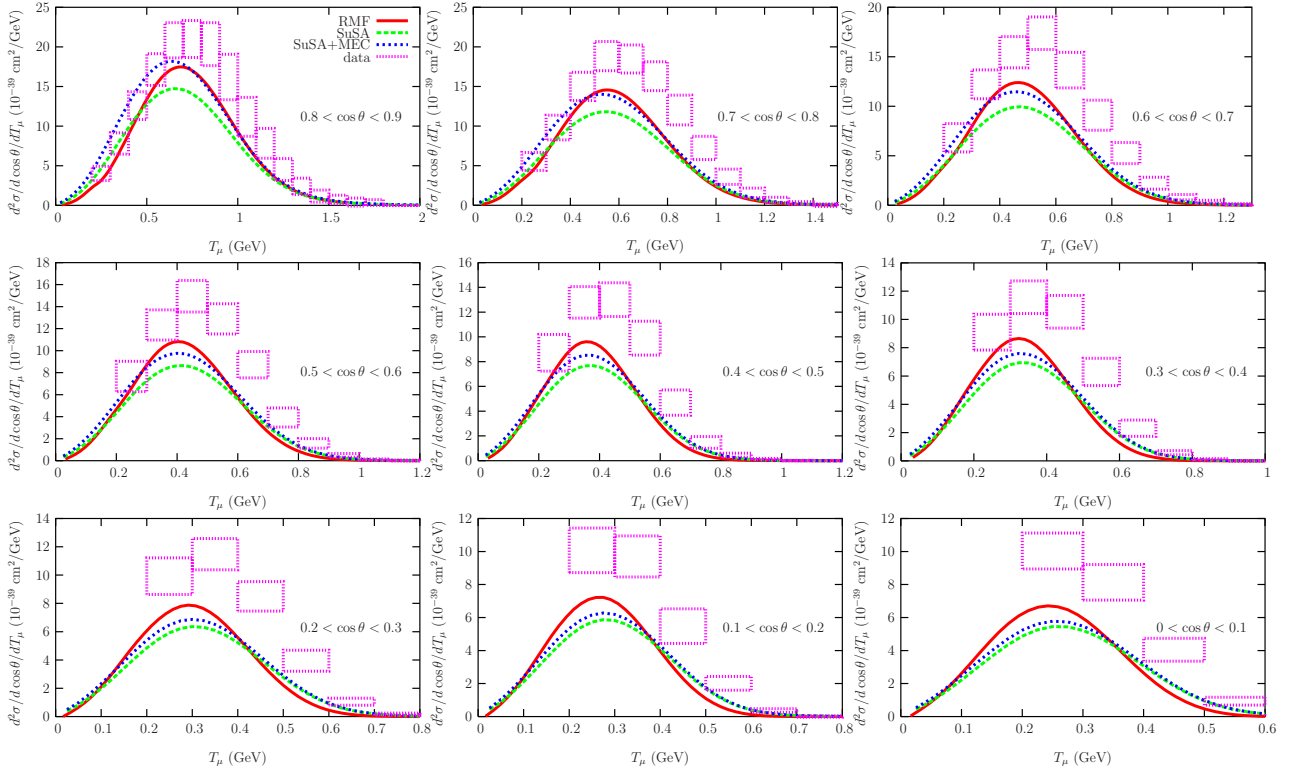

\label{fig:1}
\includegraphics[scale=0.45]{RMFvsSusa_c085.epsi}%
\includegraphics[scale=0.45]{RMFvsSusa_c075.epsi}%
\includegraphics[scale=0.45]{RMFvsSusa_c065.epsi}%
\\
\includegraphics[scale=0.45]{RMFvsSusa_c055.epsi}%
\includegraphics[scale=0.45]{RMFvsSusa_c045.epsi}%
\includegraphics[scale=0.45]{RMFvsSusa_c035.epsi}%
\\
\includegraphics[scale=0.45]{RMFvsSusa_c025.epsi}%
\includegraphics[scale=0.45]{RMFvsSusa_c015.epsi}%
\includegraphics[scale=0.45]{RMFvsSusa_c005.epsi}%
\caption{(color online) Flux-integrated double-differential cross
section per target nucleon for the $\nu_\mu$ CCQE process on
$^{12}$C evaluated in the SuSA (green line), SuSA+MEC (blue) and
RMF (red) models and displayed versus the muon kinetic energy
$T_\mu$ for various bins of $\cos\theta$. The data are from
MiniBooNE~\cite{AguilarArevalo:2010zc}. The uncertainties do not
include the overall normalization error $\delta N$=10.7\%. }
\end{figure*}

In Fig.~1 we show the double-differential cross section averaged
over the neutrino energy flux as a function of the muon kinetic
energy $T_\mu$. In each panel the results have been averaged over
the corresponding angular bin of $\cos\theta$. In all cases we use the standard value of the nucleon axial mass, {\it i.e.,}
$M_A=1.03$ GeV/c$^2$. We compare the theoretical results evaluated using the three approaches, SuSA (green line), SuSA+MEC (blue) and RMF (red), with the MiniBooNE data~\cite{AguilarArevalo:2010zc}.
The case of the most forward angles, $0.9<\cos\theta<1$, has not
been considered since, as shown in \cite{Amaro11}, models based on quasi-free scattering cannot describe properly this kinematic
situation where roughly 1/2 of the total cross section arises from excitation energies below $\sim$50 MeV.

The analysis of the results corresponding to SuSA and SuSA+MEC
approaches and their comparison with data were already presented and
discussed at length in \cite{Amaro11}. We showed that the 2p-2h MEC
increase the cross section, yielding results that are closer to
experiment, specifically, for data up to $\cos\theta\sim 0.6$. At
larger angles, the discrepancy with experiment becomes larger while,
on the other hand, the role of MEC is seen to be less significant,
that is, the difference between SuSA and SuSA+MEC becomes smaller as
the scattering angle increases.

Cross sections evaluated with the RMF model also yield reasonable
agreement with data for smaller angles, the discrepancy becoming
larger as $\theta$ increases. However, some differences emerge from
the comparison between the RMF and SuSA predictions. As observed,
RMF cross sections are in general larger than the SuSA ones. In
particular, in the region close to the peak in the cross section,
the RMF result becomes larger than the one obtained with SuSA+MEC.
This holds especially for large scattering angles. On the contrary,
SuSA and SuSA+MEC get more strength in the region of high muon
kinetic energies. This can be attributed to the breakdown of
zeroth-kind scaling in the RMF, in contrast to the other approaches
where it is assumed to be satisfied. An approach based on RMF, but
invoking zeroth-kind scaling, yields results that are much more
similar to the SuSA ones.

\begin{figure}[ht]
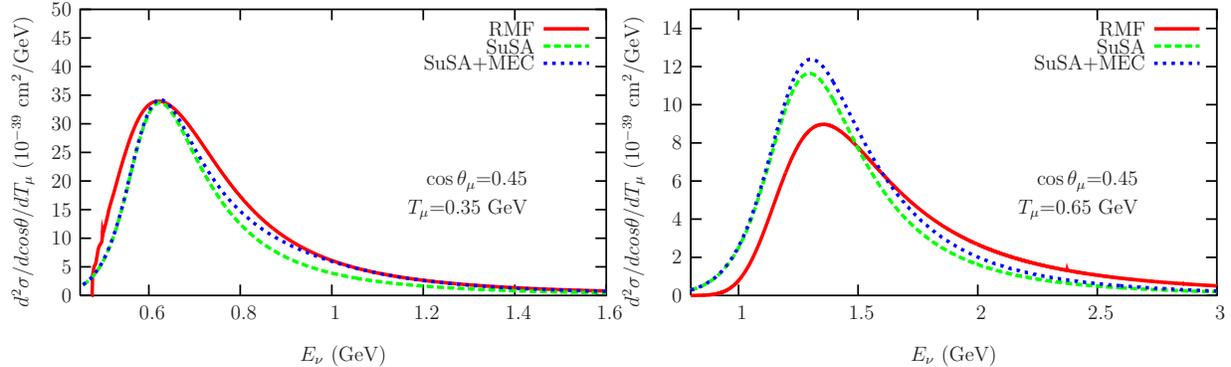

\label{fig:2}
\includegraphics[scale=0.65]{c045t035_new.epsi}%
\includegraphics[scale=0.65]{c045t065.epsi}
\caption{(color online) Double-differential cross section
calculated for fixed values of the muon kinetic energy and
scattering angle and displayed versus the neutrino energy. Results
presented for the three models: SuSA (green), SuSA+MEC (blue) and
RMF (red).}
\end{figure}

To make such a statement more transparent, we compare the
double-differential cross sections evaluated with the three
models, but for fixed values of the scattering angle and muon
kinetic energy. The results are presented against the neutrino
energy. We have selected as a representative situation the case
$\cos\theta=0.45$ (panel in the middle of Fig.~1) and two values
of $T_\mu$: 0.35 GeV that corresponds to the maximum in the
neutrino-flux-averaged cross section, and $T_\mu=0.65$ GeV,
located in the tail. Results are presented in Fig.~2. As shown,
for $T_\mu=0.35$ GeV (left panel) the three models produce roughly the same response in the maximum at $E_\nu\approx 0.6$ GeV. However, the strength in the tail for higher neutrino energies is much more significant for RMF, being reduced for SuSA+MEC and much smaller for SuSA. This explains why the RMF neutrino-flux-averaged cross section is significantly higher at $T_\mu=0.35$ GeV (see Fig.~1).

The situation is clearly different for $T_\mu=0.65$ GeV (right panel
in Fig.~2). Here, SuSA and SuSA+MEC cross sections are larger
(compared with RMF) even in the region where the cross section
reaches its maximum. On the contrary, for larger $T_\mu$ (located in
the tail), RMF becomes higher. However, notice that for these
kinematics the neutrino energies involved are much larger than in
the previous case, namely, $E_\nu\geq 0.8-0.9$ GeV; this corresponds
to the tail in the experimental neutrino flux whose average neutrino
energy is 788 MeV. Hence, the main contribution in the averaged
cross section comes from the region with smaller values of $E_\nu$
where the difference between SuSA (and SuSA+MEC) and RMF is larger.
As already mentioned, if one does an RMF calculation that respects
zeroth-kind scaling, the results would be essentially in agreement
with those of SuSA.

\begin{figure}[ht]
\label{fig:3}
\includegraphics[scale=0.85]{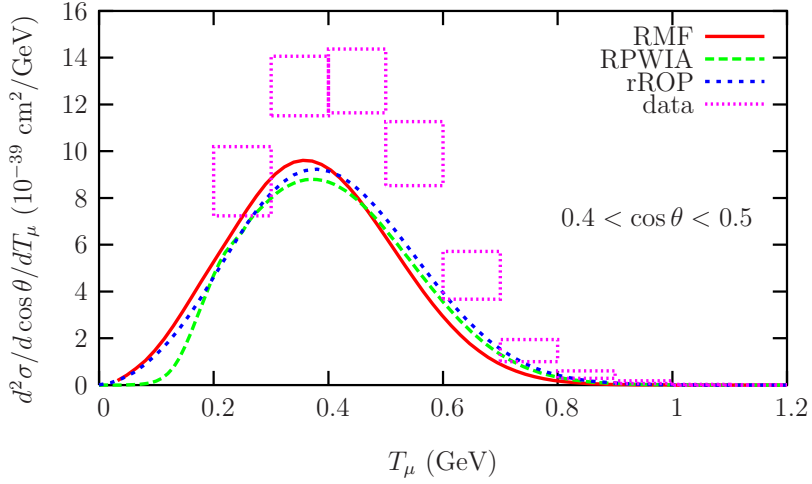}%
\caption{(color online) As for Fig.~1 for the bin
$0.4<\cos\theta<0.5$, but now showing the results evaluated with
RPWIA (green), rROP (blue) and RMF (red).}
\end{figure}
\begin{figure*}[ht]
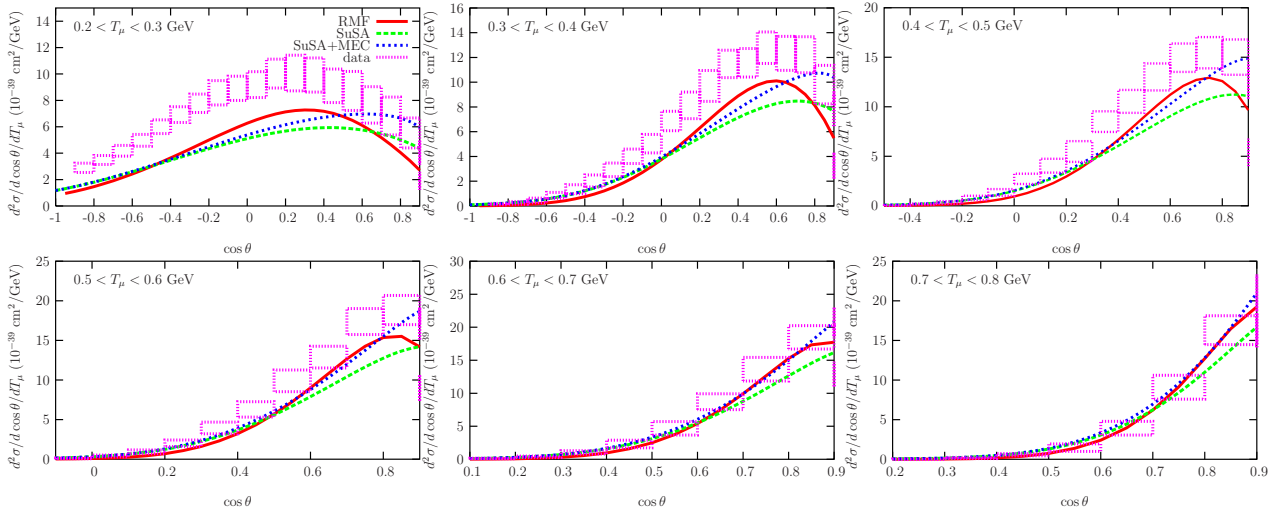

\label{fig:4}
\includegraphics[scale=0.45]{RMFvsSuSA_t025.epsi}%
\includegraphics[scale=0.45]{RMFvsSuSA_t035.epsi}%
\includegraphics[scale=0.45]{RMFvsSuSA_t045.epsi}%
\\
\includegraphics[scale=0.45]{RMFvsSuSA_t055.epsi}%
\includegraphics[scale=0.45]{RMFvsSuSA_t065.epsi}%
\includegraphics[scale=0.45]{RMFvsSuSA_t075.epsi}%
\caption{(color online) Double-differential $\nu_\mu$ CCQE cross
section for $^{12}$C integrated over neutrino flux versus the
outgoing muon scattering angle for various bins of the muon
kinetic energy $T_\mu$. Results are given for RMF (red lines),
SuSA (green) and SuSA+MEC (blue).}
\end{figure*}

For completeness in Fig.~3 we show the flux-integrated cross section
averaged over the bin $0.4<\cos\theta<0.5$ evaluated within the
framework of the relativistic impulse approximation (RIA), but with
different descriptions for the FSI. We have considered the
relativistic plane wave impulse approximation (RPWIA), that is,
switching off FSI in the RMF calculation, and the use of the real
part of the relativistic energy-dependent optical potential, denoted
as rROP. As already shown in previous
works~\cite{Caballero:2006wi,Caballero:2005sj}, these two approaches
fulfill scaling, but give rise to scaling functions that lack the
asymmetry shown by data. Moreover, scaling of the zeroth kind is
also highly respected because of the minor role played in this case
by relativistic dynamics in the final state.

Results in Fig.~3 show that the RPWIA and rROP approaches are very similar for all $T_\mu$-values, being also in accordance with RMF, although here the maximum in the cross section is slightly reduced while strength is shifted to larger values of the muon kinetic energy. This is a consequence of the differences introduced in the scaling functions by the particular description of FSI and its impact on the relativistic nuclear dynamics and the isospin (third-kind) and zeroth-kind scaling
violations~\cite{Caballero:2007tz}.

In Fig.~4 we plot the neutrino-flux-averaged cross section versus
the scattering angle at fixed $T_\mu$ (averaged over each bin). For low muon momenta the three models tend to underestimate the data, improving the agreement as $T_\mu$ increases. As observed, when added to the SuSA results, the 2p-2h MEC yield an enhancement of the cross section whose magnitude increases for more forward scattering angles. This result holds for each bin in $T_\mu$. With respect to comparison with data, some general comments already made for the SuSA and SuSA+MEC results~\cite{Amaro11} also apply to RMF: the last also underestimates the data at large muon scattering angles, particularly for small $T_\mu$. However, some important differences between RMF and SuSA-based models also emerge. Let us comment on the general trend followed by the RMF results as functions of $\cos\theta$ that clearly differ from SuSA and SuSA+MEC. In the six
panels presented in Fig.~4 RMF cross sections are the lowest for the smallest values of $\cos\theta$. As we move to more positive
$\cos\theta$, the RMF cross section grows faster, lying above the
results corresponding to SuSA+MEC in the intermediate region.
Finally, for smaller values of the scattering angle, namely
$\cos\theta$ approaching 0.9, while RMF inverts its behavior and
decreases very rapidly, SuSA and SuSA+MEC approaches to
$\cos\theta=0.9$ show a much softer slope. In fact, this is the
region where the discrepancy between RMF and SuSA-based models can be better appreciated. It is very illustrative to point out that the general shape presented by the RMF cross section as a function of $\cos\theta$ fits perfectly well the shape shown by data, although RMF predictions fall below the data for small muon momenta. The different behaviour of the models is partly due to the fact that the RMF is better describing the low-energy excitation region whereas, as already pointed out, the SuSA model has no predictive power at very low angles, where the cross section is dominated by low excitation energies and the superscaling ideas are not supposed to apply. For this reason we do not show results for the highest $\cos\theta$ bin.

\begin{figure}[ht]
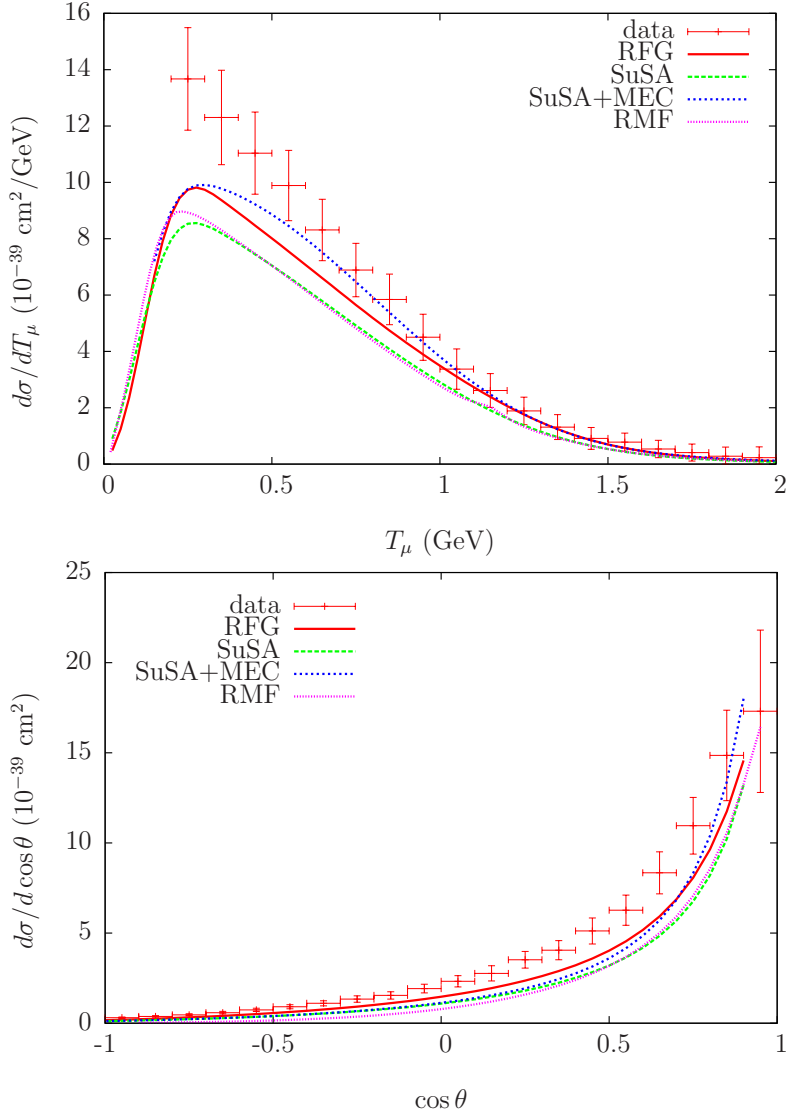

\label{fig:5}
\includegraphics[scale=0.85]{dsdt.epsi}%
\\
\includegraphics[scale=0.85]{dsdcos.epsi}
\caption{(color online) Results obtained with SuSA, SuSA+MEC, RMF
and RFG models. Upper panel: Flux-averaged integrated cross section
displayed versus the muon kinetic energy. Bottom panel: As for the
upper one, but now for the flux-averaged muon angular distribution.}
\end{figure}

In Fig.~5 we present the results obtained by integrating the
flux-averaged double-differential cross sections over $\cos\theta$
(upper panel) and $T_\mu$ (bottom panel), respectively. In addition
to the three models considered in previous graphs, here we also
include for reference the predictions given by the RFG. It is
interesting to remark that, in spite of the clear differences shown
by the RMF and SuSA predictions for the double-differential cross
sections (Figs.~1 and 4), the integrated results almost coincide. On
the contrary, the 2p-2h MEC effects produce a visible enhancement in
the cross section that is closer to the experimental data. The RFG
results lie somewhere between the SuSA/RMF and SuSA+MEC predictions.

\begin{figure}[ht]
\label{fig:6}
\includegraphics[scale=0.85]{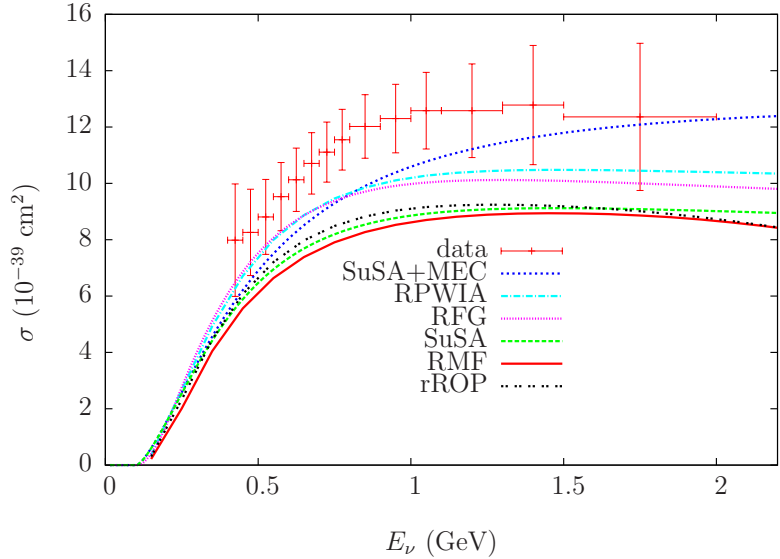}%
\caption{(color online) Total CCQE cross section per neutron
versus the neutrino energy. The curves corresponding to different
nuclear models (see text) are compared with the flux unfolded
MiniBooNE data~\cite{AguilarArevalo:2010zc}.}
\end{figure}

To conclude this section, in Fig.~6 we display the total QE cross
section per neutron obtained in the models discussed above as a
function of the neutrino energy and compared with the experimental
data. Note that here the integration is performed over all muon
scattering angles ($-1<\cos\theta<1$) and energies
$0<T_\mu<E_\nu$).

As observed in Fig.~5, the discrepancies between the various models
tend to be washed out by the integration, yielding very similar
results for the models that include FSI (SuSA, RMF and rROP), all of
them giving a lower total cross section than the models without FSI
(RFG and RPWIA). On the other hand the SuSA+MEC curve, while being
closer to the data at high neutrino energies, has a somewhat
different shape with respect to the other models, in qualitative
agreement with the relativistic calculation of \cite{Nieves:2011pp}.
It should be noted, however, that the result is affected by an
uncertainties of about 5\% associated with the treatment of the
2p-2h contribution at low momentum transfers.

\section{Concluding remarks}
\label{sec:concl}

Summarizing, in this paper we extend the previous work presented in
\cite{Amaro11} where the focus was placed on the use of the
phenomenological SuSA model and its extension incorporating the role
played by 2p-2h MEC contributions. Here, our main interest resides
in the predictions provided by the RMF approach. This model has been
successfully applied to inclusive QE $(e,e')$ processes where it has
been shown to be capable of reproducing the specific asymmetric
shape shown by the experimental scaling function. On the other hand,
and unlike the SuSA approach which assumes scaling of the zeroth
kind, {\it i.e.,} equal longitudinal and transverse scaling
functions, the RMF provides a transverse scaling function that
exceeds by about $\sim$20$\%$ the longitudinal one. This result
seems to be in accordance with the most recent analyses of the $L/T$
separated $(e,e')$ data. Thus, this violation of scaling of the
zeroth kind has visible effects when proceeding to studies of CCQE
cross sections. Furthermore, the different isospin content, (namely,
violations of third-kind scaling) of the electromagnetic and CC weak
nucleon form factors should also be carefully
considered~\cite{Caballero:2007tz}.

In this work we apply the RMF model to CCQE neutrino reactions on
$^{12}$C corresponding to the kinematics of the MiniBooNE
experiment. Results for the flux-averaged double-differential cross sections are compared with data and the predictions given by SuSA and SuSA+MEC models. Generally speaking, the RMF model
underestimates the data especially at large muon scattering angles and low muon energies. This was already observed with SuSA and to a somewhat lesser extent with SuSA+MEC. However, the specific behavior shown by RMF clearly differs from that of SuSA and SuSA+MEC; the maximum in $d^2\sigma/d\cos\theta dT_\mu$ as a function of $T_\mu$ for various bins of $\cos\theta$ gets higher for RMF, whereas the tail at high $T_\mu$ is more pronounced for the SuSA-based models. Also, the general trend shown by the curve corresponding to the double-differential cross section as a function of the scattering angle for bins of $T_\mu$, clearly differs for RMF and SuSA (SuSA+MEC) approaches. Here, it is very interesting to point out that the specific shape followed by RMF predictions fits perfectly well the slope shown by data.

The single-differential cross sections shown in Fig.~5 where the
three calculations yield very similar predictions, with almost the same shape and underpredicting the data, also shows that it is very useful to compare double-differential cross sections as in figure 4, where differences among models may be more easily seen.

To conclude, let us note that, in spite of the discrepancies
introduced by the models in the double-differential cross sections, RMF and SuSA approaches provide almost identical results for  the single-differential cross section, this being found to lie below the data. Although the inclusion of 2p-2h MEC contributions increases the differential cross section without a significant change of the shape,  and thus seems to improve the agreement with the data as shown in Figs. 2 and 5, it is also seen (in Fig. 2, but more clearly in Fig. 4) that the shape of the cross section is best reproduced by the RMF and does not improve with the inclusion of the 2p-2h MEC contributions in SuSA. It is tempting to hypothesize that addition of the 2p-2h MEC effects to the RMF results would lead to reasonable
agreement in both magnitude and shape with the experimental
double-differential cross section.

Finally, as shown in Fig.~6, the impact of the 2p-2h contribution
on the total cross section increases with the neutrino energy,
suggesting that the data can be explained without the need for a
large nucleon axial mass. However more refined calculations taking care of correlation currents, MEC effects in the axial-vector channel, {\it etc.,} should be performed before definitive
conclusions can be drawn.

\section*{Acknowledgments}
This work was
partially supported by DGI (Spain): FIS2008-01143, FPA2010-1742
FIS2008-04189, by the Junta de
Andaluc\'{\i}a, by the INFN-MEC collaboration agreement, projects
FPA2008-03770-E-INFN, ACI2009-1053, the Spanish Consolider-Ingenio
2000 programmed CPAN (CSD2007-00042), and part (TWD) by U.S.
Department of Energy under cooperative agreement DE-FC02-94ER40818.

%


\begin{thebibliography}{99}
%
\bibitem{AguilarArevalo:2010zc}
  A.~A.~Aguilar-Arevalo {\it et al.}  [MiniBooNE Collaboration],
  Phys.\ Rev.\ D {\bf 81}, 092005 (2010).
%
\bibitem{Bernard:2001rs}
  V.~Bernard, L.~Elouadrhiri and U.~G.~Meissner,
  J.\ Phys.\ G {\bf 28}, R1 (2002).
%
\bibitem{Barbaro:1996vd}
  M.~B.~Barbaro, A.~De Pace, T.~W.~Donnelly, A.~Molinari and M.~J.~Musolf,
  Phys.\ Rev.\  C {\bf 54}, 1954 (1996).
%
\bibitem{Ben10} O.~Benhar, P.~Coletti, and D.~Meloni,
Phys. Rev. Lett. {\bf 105}, 132301 (2010).
%
\bibitem{Jus10} C.~Juszczak, J.~T.~Sobczyk, and J.~Zmuda,
Phys. Rev. C {\bf 82}, 045502 (2010).
%
\bibitem{Maieron:2003df}
  C.~Maieron, M.~C.~Martinez, J.~A.~Caballero and J.~M.~Udias,
  Phys.\ Rev.\  C {\bf 68}, 048501 (2003).
%
%
\bibitem{Caballero:2006wi}
  J.~A.~Caballero,
  Phys.\ Rev.\  C {\bf 74} (2006) 015502.
%
\bibitem{Amaro:2006if}
  J.~E.~Amaro, M.~B.~Barbaro, J.~A.~Caballero, T.~W.~Donnelly and J.~M.~Udias,
  Phys.\ Rev.\  C {\bf 75}, 034613 (2007).
%
\bibitem{Caballero:2009sn}
  J.~A.~Caballero, M.~C.~Martinez, J.~L.~Herraiz and J.~M.~Udias,
  Phys.\ Lett.\  B {\bf 688}, 250 (2010).
%
\bibitem{Meucci:2009nm}
  A.~Meucci, J.~A.~Caballero, C.~Giusti, F.~D.~Pacati and J.~M.~Udias,
  Phys.\ Rev.\  C {\bf 80}, 024605 (2009).
%
\bibitem{Mar09} M.~Martini, M.~Ericson, G.~Chanfray, and J.~Marteau,
Phys. Rev. C {\bf 80}, 065501 (2009).
%
\bibitem{Mar10} M.~Martini, M.~Ericson, G.~Chanfray, and J.~Marteau,
Phys. Rev. C {\bf 81}, 045502 (2010).
%
\bibitem{Amaro:1998ta}
  J.~E.~Amaro, M.~B.~Barbaro, J.~A.~Caballero, T.~W.~Donnelly and A.~Molinari,
  Nucl.\ Phys.\  A {\bf 643}, 349 (1998)
  [arXiv:nucl-th/9806014].
%
\bibitem{Amaro:2002mj}
  J.~E.~Amaro, M.~B.~Barbaro, J.~A.~Caballero, T.~W.~Donnelly and A.~Molinari,
  Phys.\ Rept.\  {\bf 368}, 317 (2002).
%
\bibitem{Amaro11}
  J.~E.~Amaro, M.~B.~Barbaro, J.~A.~Caballero, T.~W.~Donnelly and C.~F.~Williamson,
  Phys.\ Lett.\  B {\bf 696}, 151 (2011).
%
\bibitem{Day:1990mf}
  D.~B.~Day, J.~S.~McCarthy, T.~W.~Donnelly and I.~Sick,
  Ann.\ Rev.\ Nucl.\ Part.\ Sci.\  {\bf 40}, 357 (1990).
%
\bibitem{Donnelly:1998xg}
  T.~W.~Donnelly and I.~Sick,
  Phys.\ Rev.\ Lett.\  {\bf 82}, 3212 (1999).
%
\bibitem{Donnelly:1999sw}
  T.~W.~Donnelly and I.~Sick,
  Phys.\ Rev.\  C {\bf 60}, 065502 (1999).
%
\bibitem{Maieron:2001it}
  C.~Maieron, T.~W.~Donnelly and I.~Sick,
  Phys.\ Rev.\  C {\bf 65}, 025502 (2002).
%
\bibitem{Alberico:1988bv}
  W.~M.~Alberico, A.~Molinari, T.~W.~Donnelly, E.~L.~Kronenberg and J.~W.~Van Orden,
  Phys.\ Rev.\  C {\bf 38}, 1801 (1988).
%
\bibitem{Amaro:2005dn}
  J.~E.~Amaro, M.~B.~Barbaro, J.~A.~Caballero, T.~W.~Donnelly and C.~Maieron,
  Phys.\ Rev.\  C {\bf 71}, 065501 (2005).
%
\bibitem{Amaro:2004bs}
  J.~E.~Amaro, M.~B.~Barbaro, J.~A.~Caballero, T.~W.~Donnelly, A.~Molinari and I.~Sick,
  Phys.\ Rev.\  C {\bf 71} (2005) 015501.
%
\bibitem{Vignote:2003er}
  J.~R.~Vignote, M.~C.~Martinez, J.~A.~Caballero, E.~Moya de Guerra and J.~M.~Udias,
  Phys.\ Rev.\  C {\bf 70}, 044608 (2004).
%
\bibitem{Udias:1993zs}
  J.~M.~Udias, P.~Sarriguren, E.~Moya de Guerra, E.~Garrido and J.~A.~Caballero,
  Phys.\ Rev.\  C {\bf 51}, 3246 (1995).
%
\bibitem{private11} T.~W.~Donnelly and C.~F.~Williamson (private
communication).
%
\bibitem{Caballero:2007tz}
  J.~A.~Caballero, J.~E.~Amaro, M.~B.~Barbaro, T.~W.~Donnelly and J.~M.~Udias,
  Phys.\ Lett.\  B {\bf 653}, 366 (2007).
%
\bibitem{Martinez:2005xe}
  M.~C.~Martinez, P.~Lava, N.~Jachowicz, J.~Ryckebusch, K.~Vantournhout and J.~M.~Udias,
  Phys.\ Rev.\  C {\bf 73}, 024607 (2006).
%
\bibitem{Caballero:2005sj}
  J.~A.~Caballero, J.~E.~Amaro, M.~B.~Barbaro, T.~W.~Donnelly, C.~Maieron and J.~M.~Udias,
  Phys.\ Rev.\ Lett.\  {\bf 95}, 252502 (2005).
%
\bibitem{Amaro:2006tf}
  J.~E.~Amaro, M.~B.~Barbaro, J.~A.~Caballero and T.~W.~Donnelly,
  Phys.\ Rev.\ Lett.\  {\bf 98} (2007) 242501.
%
\bibitem{HerraizZZ}
  J.~L.~Herraiz, M.~C.~Martinez, J.~M.~Udias and J.~A.~Caballero,
  Acta Phys.\ Polon.\  B {\bf 40}, 2405 (2009).
%
\bibitem{Jourdan:1996ut}
  J.~Jourdan,
  Nucl.\ Phys.\  A {\bf 603} (1996) 117.
%
\bibitem{Coulomb}
  J.~M.~Udias, P.~Sarriguren, E.~Moya de Guerra, E.~Garrido and J.~A.~Caballero,
  Phys.\ Rev.\ C {\bf 48}, 2731 (1993); Phys.\ Rev.\ C {\bf 51}, 3246 (1995).
%
\bibitem{Horowitz}
C.~J.~Horowitz, B.~D.~Serot, Nucl. Phys. A {\bf 368}, 503 (1981); Phys. Lett. B {\bf 86}, 146 (1979).
%
\bibitem{Serot}
B.~D.~Serot, J.~D.~Walecka, Adv. Nucl. Phys. {\bf 16}, 1.
Eds. J.~W.~Negele, E.~W.~Vogt. Plenum Press, New York (1986).
%
\bibitem{Sharma:1993it}
  M.~M.~Sharma, M.~A.~Nagarajan and P.~Ring,
  Phys.\ Lett.\  B {\bf 312}, 377 (1993).
%
\bibitem{Amaro:2003yd}
  J.~E.~Amaro, M.~B.~Barbaro, J.~A.~Caballero, T.~W.~Donnelly and A.~Molinari,
  Nucl.\ Phys.\  A {\bf 723}, 181 (2003).
%
\bibitem{Amaro:2009dd}
  J.~E.~Amaro, M.~B.~Barbaro, J.~A.~Caballero, T.~W.~Donnelly, C.~Maieron and J.~M.~Udias,
  Phys.\ Rev.\  C {\bf 81} (2010) 014606.
%
\bibitem{De Pace:2003xu}
  A.~De Pace, M.~Nardi, W.~M.~Alberico, T.~W.~Donnelly and A.~Molinari,
  Nucl.\ Phys.\  A {\bf 726}, 303 (2003).
%
\bibitem{Amaro:2p2h}
  J.~E.~Amaro, C.~Maieron, M.~B.~Barbaro, J.~A.~Caballero and T.~W.~Donnelly,
  Phys.\ Rev.\ C {\bf 82}, 044601 (2010).
%
\bibitem{Umino:1996cz}
  Y.~Umino and J.~M.~Udias,
  Phys.\ Rev.\  C {\bf 52}, 3399 (1995)
  [arXiv:nucl-th/9602003].
%
\bibitem{Umino:1994wu}
  Y.~Umino, J.~M.~Udias and P.~J.~Mulders,
  Phys.\ Rev.\ Lett.\  {\bf 74}, 4993 (1995).
%
\bibitem{Nieves:2011pp}
  J.~Nieves, I.~R.~Simo and M.~J.~V.~Vacas,
  arXiv:1102.2777 [hep-ph].





\end{thebibliography}
\end{document}